\documentclass[twocolumn,showpacs,preprintnumbers,amsmath,amssymb,prb]{revtex4}

\usepackage{graphicx}
\usepackage{dcolumn}
\usepackage{bm}

\begin{document}

\title{Hydrodynamic simulations of metal ablation by femtosecond laser irradiation}
\author{J.P. Colombier$^{1,2}$}
\author{P. Combis$^{1}$}%
\author{F. Bonneau$^{1}$}%
\author{R. Le Harzic$^{2}$}%
\author{E. Audouard$^{2}$}
\affiliation{$^{1}$CEA/DAM Ile de France, Dept. de Physique Th\'eorique et Appliqu\'ee, B.P. 12, 91680 Bruy\`eres-le-Ch\^{a}tel, France\\
$^{2}$Laboratoire Traitement du Signal et Instrumentation (TSI), Universit\'e Jean Monnet, UMR CNRS 5516, 42000 Saint-Etienne, France}%

\date{\today}

\begin{abstract}
Ablation of Cu and Al targets has been performed with \hbox{170 fs} laser pulses in the intensity range of \hbox{$10^{12}$-$10^{14}$ W/cm$^{2}$}. 
We compare the measured removal depth with 1D hydrodynamic simulations. The electron-ion temperature decoupling is taken into account using the standard "two-temperature model".
 The influence of the early heat transfer by electronic thermal conduction on hydrodynamic material expansion and mechanical behavior is investigated. 
A good agreement between experimental and numerical matter ablation rates shows the importance of including solid-to-vapor evolution of the metal in the current modeling of the laser matter 
interaction.
\end{abstract}



\maketitle

\section{Introduction}
Ultrafast phenomena driven by subpicosecond laser pulses have been the subject of thorough investigation for many years in order to explain the ablation of solid materials~\cite{Komashko,Laville,Schafer,Zhigilei}. As opposed to the laser-dielectric interaction where thermal and athermal ablation regimes probably takes place~\cite{Stoian,Bulgakova,Feit}, the laser-metal interaction is mainly governed by the thermal ablation one~\cite{Gamaly,Bulgakova}. The laser energy is absorbed by the free electrons first. The pulse duration being shorter than the electron-phonon relaxation time \hbox{($\tau_{e-ph} \sim 10^{-12}s$)}, electrons and ions subsystems must be considered separately. The "two-temperature model" (TTM) describes the thermal transport of energy by the electrons and the energy transfer from the electrons to the lattice~\cite{Anisimov}.

Numerous theoretical and experimental previous works have been devoted to the study of the matter ablation with a single laser pulse. Experimental irradiation conditions in current applications are largely investigated to optimize the ablation rate: pulse duration~\cite{Sallé}, fluence~\cite{Nolte} and background gas~\cite{Wynne,Preuss}. However, a complete view including all the relevant physical mechanisms is still lacking. To get a better understanding of the ablation process and to extend the results into situations not covered by the experiments, two kinds of investigations are put at work : (i) a complete identification of the various physical mechanisms responsible for the material removal from the surface, (ii) an evaluation of the impact of these various processes on the amount of ablated matter.
In the works previously addressed, few calculations are able to provide a direct comparison with experiments. Most of them are focused on thermal transport and a more detailed description of the physical processes occuring in the material has to be incorporated to really describe the whole ablation process. Among these, works based on hydrodynamic modeling~\cite{Eidmann,Laville,Afanasiev,Komashko} have been recently associated to the TTM to describe ablation process. To overcome the drawbacks of a material fluid treatment, a mechanical extension of the TTM has been proposed to model the ultrafast thermomelasticity behavior of a metal film~\cite{Chen}. Works based on molecular dynamics allow to access to the influence of the ultrafast energy deposition on the thermal and mechanical evolution properties of the material~\cite{Perez,Meunier}. With a different approach, other authors have performed a microscopic analysis of the mechanisms of ultrashort pulse laser melting by means of a hybrid molecular-dynamic and fluid modelisation~\cite{Schafer,Zhigilei}.

From all these investigations, it appears that none of these effects may be neglegted to reproduce the features of the damage resulting from an ultrashort laser irradiation. Moreover, there is a lack of investigations which combine experimental and theoretical results so that current models are still questionnable. In the present simulations, the TTM provides energy deposition in an expanding material intimately correlated with the processes governing the ablation in the ultrashort pulse case, which is a specificity of our hydrodynamic approach. Simulation results give useful insight into the presented experiment data.

Transport properties of electrons are not very well understood in nonequilibrium electron-ion systems. However, the comprehension of these phenomena in the context of ultrafast interaction is essential not only for fundamental purposes but also for micromachining applications. A precise description of the effect of the electronic temperature on the absorption seems to be still unsettled~\cite{Milchberg,Fisher,Rethfeld}, and it has not been taken into account in the presented calculations. Nevertheless, the model employed in this work uses a large set of current available data.
\newline \indent
Obviously, numerical calculations are always requiring additional information such as electron-phonon or electron-electron relaxation times, which may be extracted, from experimental data~\cite{Corkum,Gusev}. Reciprocally, comparison between simulations and experiments allows to validate physical data introduced into the theoretical models. For instance, we shall see in the following that the measure of the pressure variation with time inside the material would be very informative. These data, however, are difficult to measure with a high accuracy. By contrast, experimental measure of laser ablation rate seems to be easier to be obtained and compared with numerical simulations.
\newline \indent
In this article, numerical and experimental results on ablated matter are reported. For this purpose, the TTM is inserted into a hydrodynamic code in order to describe the material removal. First we detail the physical processes which are taken into account in our computations within the framework of hydrodynamical modeling. Then, the effects caused by relaxation processes on the evolution of shock waves are examined. We next present the experiments which have been performed to obtain ablation depth measurements as a function of the laser fluence. Finally, our discussions are based on results of numerical simulations on Cu and Al samples compared with specific experimental measurements.

\section{Theoretical model}
\begin{figure}[htbp]
\includegraphics[width=8.5cm]{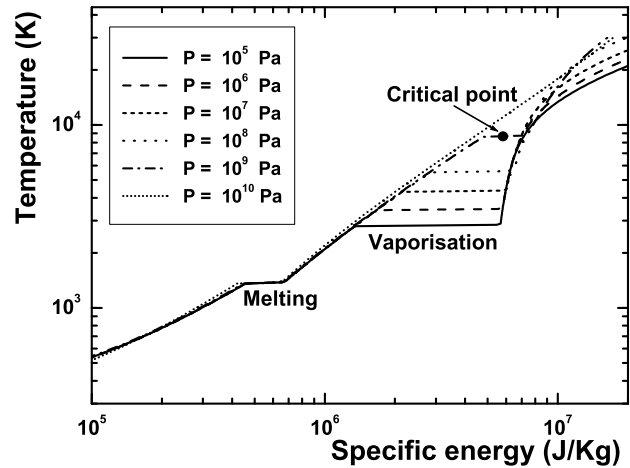}
\caption{Representation of isobars in a phase diagram of the copper EOS in the region of phase transitions.\label{EOS_cu}}
\end{figure}

To represent numerically the interaction between the laser and the metallic target, we used a 1D Lagrangian hydrodynamic code~\cite{Delpor}. Solving the Helmholtz wave equation permitted us to determine the electromagnetic field through the region illuminated by the laser. The deposited energy is then deduced using the Joule-Lenz law. 
A precise result for the absorption from an arbitrary medium can be obtained from the direct solution of the equation for the electromagnetic field. Let us consider a planar wave propagating along z axis. We write the 
following Helmholtz equation for the complex amplitude of the electric field $E_{z}$ with frequency $\omega$ :
\begin{equation}\label{three}
\Delta E_{z}-\nabla\nabla E_{z}+\left(1+i\frac{4\pi}{\omega}\sigma\left(T,\rho\right)\right)\frac{\omega^{2}}{c^{2}}E_{z}=0  
\end{equation}
where $\sigma(T,\rho)$ is the complex conductivity and $c$ is the light speed. The function $\sigma$ is calculated as a function of the density $\rho(z)$ and the temperature $T(z)$. The relative permittivity of the medium is supposed to be equal to the unity in the case of a metal target.
These simulations need accurate data 
such as transport coefficients in solids, liquids, vapors and dense or diluted plasmas, specially refractive indices~\cite{Palik}, 
electric and thermal conductivities~\cite{Ebeling}, and mechanical properties such as material strength.
The evolution of the irradiated target is governed by the fluid hydrodynamic and heat diffusion equations connected with a multi-phase equation of state (EOS). Thermodynamic functions that realistically describe characteristics of metal properties in various parts of phase diagram are needed. A such set of different metal EOS is generated by means of a numerical tool developped by the authors of the reference~\cite{EOS}. As an illustration of the EOS used, \hbox{Fig.~\ref{EOS_cu}} displays  isobars in the phase diagram temperature-specific energy of the copper for a wide range of pressure. Such data reveal the dependence on the thermodynamic properties of the melting and vaporisation processes.

The mechanical propagation of shock waves and fracture are also simulated. Elastoplasticity is described by a strain rate independent model (relevant to high strain rate conditions at high pressure, typically beyond \hbox{10 GPa}) which accounts for pressure, temperature and strain dependent yield strength and shear modulus~\cite{SCG}. 
Laser induced stresses are the combination of the hydrostatic pressure and the response to the shearing deformation. 
In the temperature range of interest here, the effect of radiative energy transport on the hydrodynamic motion is negligible. Hence we ignore energy transport by radiation when solving the hydrodynamic equations.

Ultrashort laser irradiation and the associated ionic and electronic temperature decoupling require to introduce specific electronic parameters. 
We assume that free electrons form a thermal distribution during the interaction and use the Fermi-Dirac distribution to determine the electron properties (energy, pressure and heat capacity) as a 
function of the density and the temperature~\cite{Ashcroft}. 
The evanescent electromagnetic wave is absorbed by the electrons. In our range of intensity, 
they are quickly heated at a temperature of few eV. 
During and after the pulse, the energy diffuses among the electrons and is then transferred to the lattice. 
Classical heat diffusion plays a significant role as soon as a thermal gradient occurs in the material.
Diffusion processes are described by the following equations :
\begin{eqnarray}
\label{one}
\rho \ C_{e}
\displaystyle\frac{\partial T_{e}}{\partial t}&=&
\displaystyle\nabla
\displaystyle\left(K_{e}\displaystyle\nabla
T_{e}\displaystyle\right)- \gamma (T_{e}-T_{i}) + S \hspace{1cm} \;
\\ \label{two}
\rho \ C_{i}
\displaystyle\frac{\partial T_{i}}{\partial t}&=&
\displaystyle\nabla \displaystyle\left(
K_{i}
\displaystyle\nabla
T_{i}\displaystyle\right)+ \gamma (T_{e}-T_{i})
\end{eqnarray}
where $T$, $C$ and $K$ are the temperature, the specific heat and the thermal conductivity respectively. Indices "e" and "i" stand for electron and ion species. $\rho$ is the mass density. $\gamma$ characterizes the rate of energy exchange between the two subsystems and $S$ is the space and time dependent laser source term determined by 
the Joule-Lenz law. Introduction of the TTM in a hydro-code allows us to take into account the density dependence of both specific heats and conductivities. $C_{e}(T_{e},\rho)$ is calculated with the Fermi model using $\rho$ dependent chemical potentials~\cite{Ashcroft}. The electron thermal conductivity $K_{e}$ is commonly expressed in the form~\cite{Anis-Reth}:
\begin{equation}\label{three}K_{e} = \alpha \frac{(\theta_{e}^{2}+0.16)^{5/4}(\theta_{e}^{2}+0.44)}
{(\theta_{e}^{2}+0.092)^{1/2}(\theta_{e}^{2}+\beta \theta_{i})} \theta_{e}
\end{equation}where $\theta_{e}$ and $\theta_{i}$ are electron and ion temperature normalized to the Fermi temperature ($\theta_{e}=T_{e}/T_{F}$, $\theta_{i}=T_{i}/T_{F}$) and $\alpha$, $\beta$ are material dependent parameters~\cite{Schafer,Wang}. 
The linear variation of coupling term with ($T_{e}-T_{i}$) is classic in TTM : 
we have taken $\gamma=\gamma_{0}$ as \hbox{$3\times10^{16}$ WK$^{-1} $m$^{-3}$} for copper~\cite{Corkum}, 
and \hbox{$3\times10^{17}$ WK$^{-1}$ m$^{-3}$} for aluminum~\cite{Fisher}. 
It must be noticed that the values of $K_{e}$ and $\gamma$ are subject to 
considerable uncertainty in literature~\cite{Schafer}. To accurately describe the ultrafast response, 
we incorporate electronic pressure into the set of hydrodynamic equations. The system of equations for electron and ion subsystems can be written in the Lagrangian form :
\begin{equation} \label{four}
\begin{cases}
\vspace{0.2cm}
\displaystyle\frac{\partial\varepsilon_{e}}{\partial t}=
-p_{e}\displaystyle\frac{\partial V}{\partial t}, \hspace{1.9cm}
\displaystyle\frac{\partial\varepsilon_{i}}{\partial t}=
-p_{i}\displaystyle\frac{\partial V}{\partial t}\\ 

\displaystyle\frac{\partial u}{\partial t}=
-\frac{\partial}{\partial m}(p_{e}+p_{i}),\hspace{1cm}
 \displaystyle\frac{\partial V}{\partial t}=\frac{\partial u}{\partial m}
\end{cases}
\end{equation} 
where $u$ is the fluid velocity, $m$ the mass and $V$ the specific volume. $p_{e}$, $\varepsilon_{e}$, $p_{i}$, $\varepsilon_{i}$ are the pressure and the specific energy of electrons and ions.
\begin{figure}[htbp]
\includegraphics[width=8.5cm]{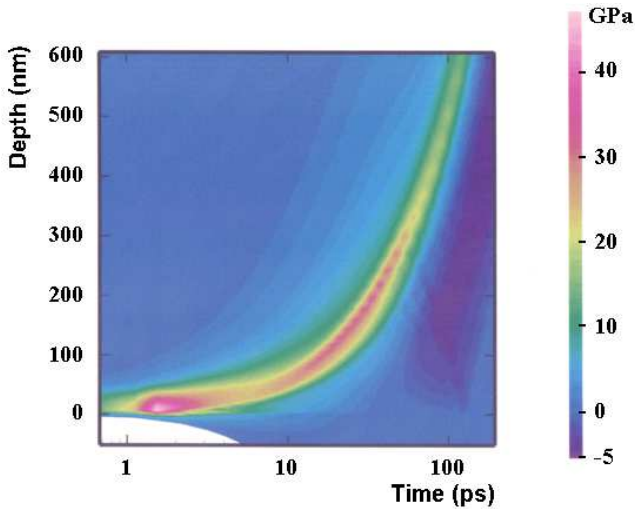}
\caption{Contour plots of the stress resulting from the irradiation of a copper target by a 
\hbox{170 fs}, \hbox{10 J/cm$^{2}$} laser pulse.\label{sptime}}
\end{figure}
 Equations~(\ref{one}) to~(\ref{four}) connected with a multiphase equation of state (EOS)~\cite{EOS} constitute a self-consistent description of the matter evolution. However, the pressure \hbox{$p(T_{e} = T_{i})$} provided by this EOS is the sum of the electronic and ionic pressure at the equilibrium and has to be replaced by the sum of these two contributions out of equilibrium.
The electronic pressure is independently determined by means of the standard fermion gas model. 
As a consequence, the total pressure used in the 
above calculations is determined as \hbox{$p(T_{e} \neq T_{i})=p_{e}(T_{e})-p_{e}(T_{i})+p_{i}(T_{i})$}.

\section{Simulations and analysis}
\begin{figure}[htbp]
\includegraphics[width=8.5cm]{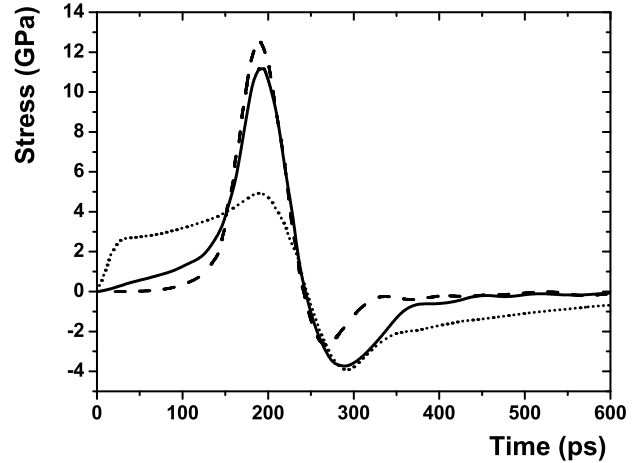}
\caption{Time evolution of the stress in copper at \hbox{1 $\mu$m} depth resulting from a \hbox{170 fs}, \hbox{10 J/cm$^{2}$} 
laser pulse. Standard conditions~:~\hbox{$\gamma = \gamma_{0}$} and \hbox{$K_{e} = K_{e_{0}}$} (solid line), \hbox{$\gamma =$5 $\gamma_{0}$} (dashed line), 
\hbox{$K_{e}$ = 10 $K_{e_{0}}$} (dotted line).\label{stress}}
\end{figure}

To start with, the interaction of a \hbox{10 J/cm$^{2}$}, \hbox{170 fs} FWHM gaussian pulse at \hbox{800 nm} wavelength 
with a copper target is investigated. \hbox{Fig.~\ref{sptime}} shows the space-time evolution of the induced stress. 
The metal surface is heated to a maximum of \hbox{2950 K}, \hbox{30 ps} after the irradiation. At this time, the free surface
 expands in a liquid state with a velocity of \hbox{400 ms$^{-1}$}. Due to the electron heating, an electronic compression wave appears at the end of the laser pulse. The electron-ion energy exchange 
results in a significant increase in the ionic pressure, which propagates inside the metal. Behind the shock, 
tensile stress occurs associated with very high strain rate around \hbox{$10^{9}$ s$^{-1}$}. \newline \indent
In order to study the sensitivity of the above results with respect to the physical parameters, we compare in \hbox{Fig.~\ref{stress}} the time variation of the computed stress at \hbox{1 $\mu$m} depth under standard conditions ($\gamma = \gamma_{0}$ and $K_{e} = K_{e_{o}}$) with those obtained with an increased coupling factor or electronic conductivity.

In the first hundred picoseconds, the stress growth is directly related to the heating depth. The increased coupling factor leads to a steeper thermal gradient and a lower temperature at \hbox{1 $\mu$m depth} compared to the other situations. The resulting stress is therefore lower in this case. The peak of the shock wave, propagating from the surface, reaches the \hbox{1 $\mu$m} depth at \hbox{200 ps}. Increasing the coupling factor accelerates the energy transfer from the electrons to the lattice and results in higher compression. Inversely, an enhanced electronic conductivity spreads the energy spatial profile and yields 
reduced stresses. It must be noticed that in the three cases, the compression is followed by a high tensile stress greater than the characteristic tensile strength of the material. Nevertheless, the loading in tension is not long enough (\hbox{150 ps}) to induce 
a fracture in the medium~\cite{Tuler-Butcher,strength}. The shock duration and intensity provide a good signature of the balance between the electronic diffusion and the electron-ion coupling. Further improvements will be discussed in the following.
\newline \indent
To obtain local information on the energy transfer induced by a 
femtosecond laser irradiation, a high-resolution time measurement of the stress reaching the rear  side of a thin sample could be 
achieved~\cite{Tollier,Romain}. Such experimental records could be directly compared with our simulations which would led us to refine the \hbox{($\gamma$, $K_{e}$)} physical values accordingly. To validate the computation, we performed ablation measurements 
and compared them to the current simulations using standard values of \hbox{($\gamma$, $K_{e}$)} for aluminum and copper samples.

\section{Experiment details}

\begin{figure}[htbp]
\includegraphics[width=8.5cm]{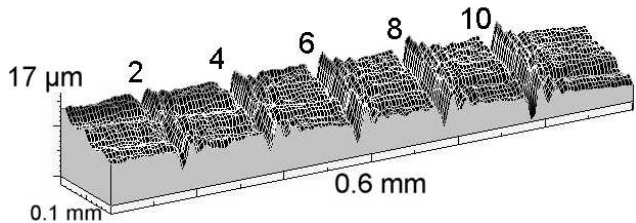}
\caption{SEM picture of machined-groove profiles, from 2 to 10 passes, on a copper sample.\label{profilo}}
\end{figure}

Ultrashort laser pulses are generated by an amplified all solid-state Ti:Saphire laser chain. Low energy pulses are extracted from a mode-locked oscillator (\hbox{1.6 nJ/pulse}, \hbox{80 MHz}, \hbox{800 nm}, \hbox{120 fs}). The pulses are then injected into an amplifying chain including~:~an optical pulse stretcher, a regenerative amplifier associated with a two-pass amplifier using a \hbox{20 W Nd:YLF laser} as pumping source, and a pulse compressor. Linearly polarized pulses with wavelength centered around 800 nm, an energy of \hbox{1.5 mJ} at \hbox{1 kHz} repetition rate and typical duration of \hbox{170 fs} were delivered. The samples are mounted on a three-motorized-axes system with 0.5 $\mu$m accuracy. Experiments are performed in the image plane of an aperture placed before the objective. The resulting spatial laser profile on sample is "top hat" so that borders and spurious conical drilling effects are reduced. Focusing objectives of \hbox{25 mm} or \hbox{10 mm} focal lengths to obtain fluences in a range of 0.5 to 35 J/cm$^{2}$ with the same beam size, 16 $\mu$m in diameter, in the image plane. 
\newline \indent
For ablation depth measurements, grooves are machined by moving the sample~\cite{these}. The sample speed is adapted such that each point on the groove axis undergoes 8 consecutive irradiations at each target pass. The number of times the sample passes in front of the fixed beam can be adjusted. 
\hbox{Fig.~\ref{profilo}} shows a scanning electron microscopy (SEM) picture of the machined grooves on copper for 2, 4, 6, 8 and 10 passes.
The groove depth is measured using an optical profilometer with a \hbox{10 nm} depth resolution. The ablation rates for each groove are deduced taking into account the sample speed,
 the repetition rate of the laser, the beam size and the number of passes. For each energy density, an averaged ablation rate is determined and the number of passes has 
been chosen to obtain reproducible results. From these experiments on copper and aluminum targets, we evaluate, for different fluences, an ablation depth averaged over a few tens of laser shots. The theoretical ablation depth is deduced from 1D numerical simulations using a criterion 
discussed hereafter.

\section{Results and discussion}

\begin{figure}[htbp]
\includegraphics[width=8.5cm]{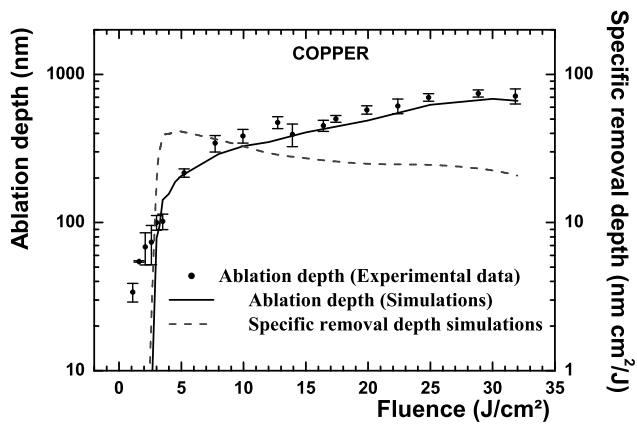}
\caption{Experimental and numerical (solid line) ablation depth as a function of the laser fluence on a copper target obtained with a 170 fs laser pulse. The dashed line shows evolution of the specific removal rate.\label{cu}}
\end{figure}

\begin{figure}[htbp]
\includegraphics[width=8.5cm]{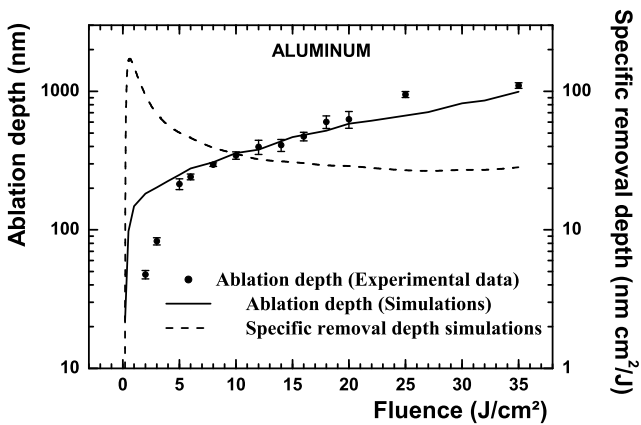}
\caption{Experimental and numerical (solid line) ablation depth as a function of the laser fluence on an aluminum target obtained with a 170 fs laser pulse. The dashed line shows evolution of the specific removal rate.\label{alu}}
\end{figure}

At moderate fluence, the propagation of the shock wave induced by the energy deposition on the lattice causes the surface 
expansion at very high speed and significant non-uniform strain rates. Simultaneously, the surface of the target is melted or
 vaporized as soon as the conditions of temperature and density required are fulfilled. High strain rates can turn the liquid region into an ensemble of droplets and ablation follows. This process is called the homogeneous nucleation~\cite{nucleation}. Unfortunately, quantitative values on the formation and ejection of liquid droplets are difficult to access because the interfacial solid-liquid microscopic properties of the nucleation centers are not accurately known. Nevertheless, in our simulation we can consider that the liquid layer accelerated outside the target corresponds to the ablated matter. Large values of strain rates (\hbox{$10^{9}$ s$^{-1}$}) indeed signal that droplet formation may occur and are sufficient to produce ablation. At higher fluence, the surface is strongly vaporized. The gas expands near the free surface and compresses the internal matter. The vaporized part of the target added to the fraction of the above-defined liquid layer constitute the calculated ablated matter. Experimental results on ablation of copper and aluminum are compared with the numerical estimates in \hbox{Fig.~\ref{cu} and~\ref{alu}}.

Sharp numerical ablation thresholds are visible at 3 and \hbox{0.5 J/cm$^{2}$} for Cu and Al targets respectively. At high fluence, the ablation saturates for both materials. This saturation occurs mainly for two reasons. Vaporization and subsequent gas expansion consume energy that does not contribute to the ablation process. Moreover, the liquid layer confinement increases as far as the gas expands and limits the liquid removal. 

As defined by Feit \emph{et al}~\cite{Feit}, a specific removal depth, i.e., depth removed per unit incident fluence could be a relevant parameter to estimate the ablation efficiency at a fixed pulse duration. Calculations of this quantity is carried out as a function of the laser fluence. Dashed curve presented in \hbox{Fig.~\ref{alu}} indicate a maximum value of \hbox{0.5 J/cm$^{2}$} in the aluminum case. The curve is smoother for copper in \hbox{Fig.~\ref{cu}} and the maximum specific removal depth is reached at a fluence around \hbox{5 J/cm$^{2}$}. This point corresponds to the occurence of a critical behavior which confirms a change in the hydrodynamic behavior. While the thickness of the liquid layer grows with the incident energy, the specific removal depth rises. Afterwards, when the gas expansion starts, a growing part of the laser energy is transformed in kinetic energy and the specific removal depth drops. This suggests that an optimum material removal exists and refers to the situation when the surface vaporisation is limited. It appears that this quantity is relevant for material processing which can be optimized by operating at this optimal fluence.

At low fluence, a noticeable discrepancy arises between the experimental and numerical results. The calculated ablated matter for a copper target is below the experimental results, while for an aluminum target, numerical results are above. We suspect that electron transport properties should be further improved. It has been shown that a significant decrease in the electrical conductivity may take place as a result of the electronic temperature increase~\cite{Milchberg}, which our model discards. 
However, the experimental measurements and the theoretical calculations come to a reasonable agreement at higher fluence. 
As it has  been shown by Fisher \emph{et al}~\cite{Fisher}, in the vicinity of a 800 nm wavelength, the interband
absorption occuring in an aluminum target decreases with increasing temperature. The authors show that, with 50 fs laser pulses, the absorbtion coefficient presents a minimum near ablation threshold, at \hbox{$5 \times 10^{13}$  W/cm$^{2}$} laser intensity. The evolution of interband absorption with the temperature is not taken into account in our calculations. Consequently, we may overestimate the energy absorbed in this intensity region.

\begin{figure}[htbp]
\includegraphics[width=8.5cm]{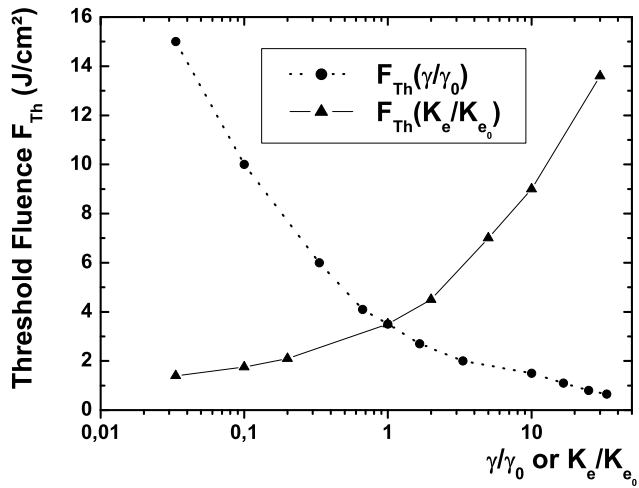}
\caption{Calculation of the threshold fluence dependence on the coupling parameter $\gamma$ and the electronic conductivity $K_{e}$ for copper. Simulations are performed for different ratio of the standard value of one parameter remaining the second one constant. Consequently, the intersection of curves coincide with the value used in the calculation of ablation rate presented in \hbox{Fig.~\ref{cu}}.\label{seuil}}
\end{figure}

For copper, the simulation overestimates the ablation threshold. This can be due to a deficit of physical process comprehension or to the inaccuracy of the parameters introduced in the model. No single value of $\gamma$ or $K_{e}$ can perfectly fit the sets of data shown in Fig.~\ref{cu} and~\ref{alu}. As for the discussion of pressure presented above, one can think that a change in $\gamma$ or $K_{e}$ has a similar effect on the threshold fluence $F_{Th}$. Therefore, it is interesting to investigate the fluence threshold dependence on $\gamma$ and $K_{e}$. A parametric analysis has been performed for a copper target. \hbox{Fig.~\ref{seuil}} displays the threshold fluence which has been obtained for different parameter couples  \hbox{($\gamma$, $K_{e}=K_{e_{0}}$)} and  \hbox{($\gamma=\gamma_{0}$, $K_{e}$)}. The deposited energy with lower thermal conductivity or higher coupling factor can penetrate deeper into the material. As a consequence, $F_{Th}$ is lowered with respect to that of the reference case  \hbox{($\gamma_{0}, K_{e_{0}}$)} and would be comparable with experimental data in the vicinity of the threshold. However, simulations performed in these conditions have shown a higher disagreement with experimental data at higher fluence due to an earlier expanded vapor. On the contrary, for a reduced $\gamma$ or an enhanced $K_{e}$ value, $F_{Th}$ increased and the ablation depth at high fluence is overestimated.

Results obtained with one-temperature simulations evidence the importance of the TTM to reproduce the experimental results. The good agreement obtained between experimental data and simulations underlines the need of coupling the TTM model with a hydrodynamic code for ablation simulations in metals. Numerical results presented in this paper give an overall description of processes occuring during ultrashort laser ablation experiments.

\section{Conclusion}
In this paper, we have reported new results on the interaction of femtosecond laser pulses with metal targets at intensities up to $10^{14}$ W/cm$^{2}$. Numerical computations were carried out by means of a 1D hydrodynamic code describing the laser field absorption and the subsequent phase transitions of matter.
 Simulations were compared to original ablation experiments performed on aluminum and copper samples. The behavior of the ablation depth as a function 
of laser fluence confirms the importance of the use of specific two-temperature equation of state and hydrodynamics. An optimum condition for laser fluence has been identified and could provide a precious information for efficient material processing.
We have highlighted that ablation process is not only governed by electronic diffusion but also by the high shock formation and propagation. The differences between experimental and numerical results remain, however, more pronounced for low laser fluences. 

We took a great care to extend the metal properties to the nonequilibrium case. Nevertheless, inclusion of realistic material parameters, such as sophisticated band structure of metals or various scattering mechanisms, would allow calculations with more accuracy. Also, a full nonequilibrium treatment should take into account the conductivity dependence with both electron-electron and electron-phonon collisions. This work is in progress and implies an elaborate optical absorption model more suitable for ultrashort laser pulse.

Simulations suggest that the in-depth stresses induced by an ultrashort laser pulse provide information of the matter dynamics in time, with which experimental pressure measurements could be compared. In particular, because it develops over temporal scales larger than the energy deposition one, the characteristic shape of the delayed shock conveys information about the interaction physics and it should thus supply a promising way for exploring matter distortions upon picosecond time scales.

\end{document}